\documentclass[journal]{IEEEtran}
\usepackage[latin1]{inputenc} 
\usepackage{graphicx}
\usepackage{float}
\usepackage[justification=centering]{caption}
\usepackage{subfigure}
\usepackage{latexsym}
\usepackage{amsmath}
\usepackage{amsthm}
\usepackage{makecell,rotating,multirow,diagbox}
\usepackage{amsfonts}
\usepackage{dsfont}
\usepackage{threeparttable}
\usepackage{multirow}
\usepackage{enumerate}
\usepackage{epsfig}
\usepackage{bm}
\usepackage{cite}
\usepackage{textcomp}
\usepackage{xcolor}
\usepackage{setspace}
\usepackage{stfloats}
\usepackage{soul, color, xcolor}
\soulregister{\cite}7
\soulregister{\ref}7

\usepackage{amssymb}
\usepackage{amsmath}
\usepackage{tabu}
\usepackage{makecell}
\usepackage[ruled,norelsize,vlined,linesnumbered]{algorithm2e}


\begin{document}
\pdfoutput=1
\title{LoS Sensing-based Channel Estimation in UAV-Assisted OFDM Systems}
\author{{Chaojin~Qing,~\IEEEmembership{Member,~IEEE,}
        ~Zhiying Liu,~Wenquan Hu,~Yinjie Zhang,~Xi Cai,~and Pengfei Du}
\thanks{This work is supported in part by the National Natural Science Foundation of China (Grant No. 62301447), the Sichuan Science and Technology Program (Grant No. 2023YFG0316, 23GSC00004), the Key Scientific Research Fund of Xihua University (Grant No. Z1320929), and the Industry-University Research Innovation Fund of China University (Grant No. 2021ITA10016).}

\thanks{C. Qing, Z. Liu, W. Hu, Y. Zhang, X. Cai, and P. Du are with the School of Electrical Engineering and Electronic Information, Xihua University, Chengdu, 610039, China (E-mail: qingchj@mail.xhu.edu.cn). }
}

\markboth{IEEE WIRELESS COMMUNICATIONS LETTERS,~Vol.~XX, No.~XX, XXX~2023}%
 {Shell \MakeLowercase{\textit{et al.}}: Bare Demo of IEEEtran.cls for IEEE Journals}


\maketitle

\begin{abstract}
In unmanned aerial vehicle (UAV)-assisted orthogonal frequency division multiplexing (OFDM) systems, the potential advantage of the line-of-sight (LoS) path, characterized by its high probability of existence, has not been fully harnessed, thereby impeding the improvement of channel estimation (CE) accuracy. Inspired by the ideas of integrated sensing and communication (ISAC), this letter develops a LoS sensing method aimed at detecting the presence of LoS path. Leveraging the prior information obtained from LoS path detection, the detection thresholds for resolvable paths are proposed for LoS and Non-LoS (NLoS) scenarios, respectively. By employing these specifically designed detection thresholds, denoising processing is applied to classical least square (LS) CE, thereby improving the CE accuracy. Simulation results validate the effectiveness of the proposed method in enhancing CE accuracy and demonstrate its robustness against parameter variations.
\end{abstract}


\section{Introduction}

Unmanned aerial vehicle (UAV)-assisted orthogonal frequency division multiplexing (OFDM) systems have garnered widespread attention for their ability to maintain reliable communication in the presence of a significant amount of ground scatter\cite{add1}\cite{r17}.
In UAV-assisted OFDM systems, channel estimation (CE) plays a crucial role in ground base station (gBS) receiver design, propelling numerous research efforts \cite{r19,r20,r21}. Although valuable insights of CE are offered in \cite{r19,r20,r21}, the distinctive characteristics of UAV-assisted communication scenarios, particularly the high probability of line-of-sight (LoS) path existence, have not been fully explored and utilized.

To comprehensively develop and leverage the advantages of LoS features, LoS sensing technology has significantly advanced in recent years \cite{r7,r31,r32}.
In UAV-assisted wireless communication systems, LoS sensing aligns with the groundbreaking idea of integrated sensing and communication (ISAC), where the received communication signals are simultaneously utilized for sensing purposes\cite{r29}.
Preliminary endeavors have been explored LoS features for channel state information feedback in UAV-assisted millimeter wave systems \cite{r30}.
However, to the best of our knowledge, few works have investigated LoS sensing-based CE in UAV-assisted OFDM systems.

Inspired by communication-centric ISAC ideas, a LoS sensing-based CE in UAV-assisted OFDM systems is proposed in this letter \footnote{
The OFDM technology serves as an example in this letter, the proposed method can also be applied to UAV-assisted wireless communication systems that do not utilize OFDM technology.}. Specifically, the LoS sensing method is utilized to detect the presence of LoS path. Leveraging the identified LoS/Non-LoS (NLoS) scenarios and motivated by the threshold superiority in ISAC scenarios\cite{add2}, we design the detection thresholds for resolvable paths in LoS and NLoS scenarios, respectively. The specialized detection thresholds eliminate false paths estimated by the classic CE method (e.g., least square (LS)) through threshold-based denoising processing, thereby improving the estimation accuracy. Our simulation results demonstrate that the proposed method is effective to improve the CE accuracy, and it is robust against the impacts of parameter variations. The main contributions of this letter are as follows:

\begin{enumerate}
    \item We explore LoS sensing-derived prior information to enhance the CE accuracy in UAV-assisted OFDM systems. This prior information is established by applying LoS/NLoS scenario sensing to capture the distinctive scenario features inherent in UAV-assisted wireless systems. Leveraging the information from received communication signals, this strategic sensing operation precedes the actual CE, thereby furnishing valuable prior insights that significantly contribute to the efficacy of the estimation process.
    \item We design LoS sensing-based detection thresholds for CE denoising. Leveraging the insights from sensing priors, the specialized detection thresholds are designed for resolvable paths in LoS and NLoS scenarios, respectively. These thresholds prove effective in mitigating the impact of false paths induced by noise or interference, thereby improving CE denoising compared to the case lacking prior information from LoS sensing.
    \item We develop a LoS sensing-based CE method in UAV-assisted OFDM systems. By harnessing insights from sensing priors, the proposed method significantly enhances the CE accuracy in comparison to existing CE methods. Additionally, our approach exhibits robustness against parameter variations. Furthermore, the proposed method provides a systematic methodology for sensing-assisted CE in operational $4$G/$5$G systems adopting OFDM technology.
\end{enumerate}

The rest of this letter is organized as follows. Section II illustrates the system model of UAV-assisted OFDM. In Section III, an enhanced CE method based on LoS sensing is proposed. Section IV contains the simulation results and discussion. Finally, we conclude the letter in Section V.

\textit{Notations}:
Bold face lower case and upper case letters represent vector and matrix, respectively. ${\left(\cdot \right)^\mathrm{T}}$ and ${\left(\cdot \right)^\mathrm{H}}$ denote the transpose and conjugate transpose, respectively. ${{\bf{F}}_N}$ is the normalized $N \times N $ Fourier transform matrix, $\mathbf{I}_N$ stands the $N$-order unit matrix, and $E\left[ \cdot \right ]$ denotes expectations.

\section{System Model}

\begin{figure}[htbp]
        \centering
        \includegraphics[scale=0.5]{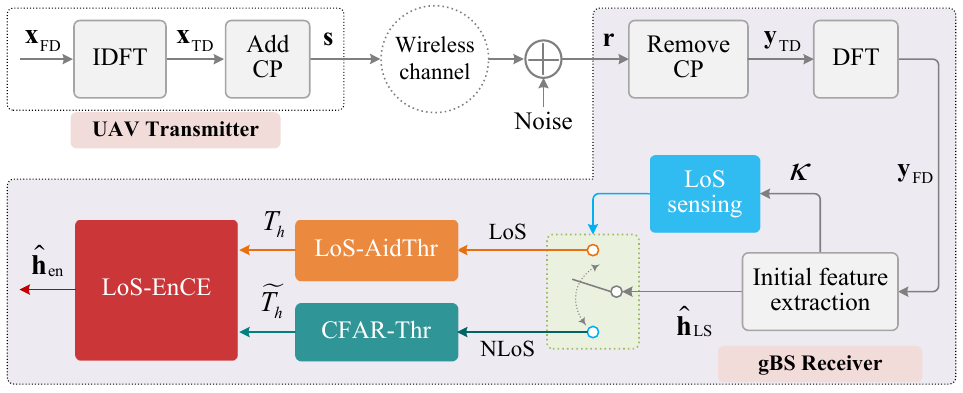}
        \captionsetup{font={footnotesize}}
        \caption{Architecture of the proposed joint processing system.}
        \label{figure1}
    \end{figure}

The diagram of the UAV-assisted wireless systems is illustrated in Fig.~\ref{figure1}, where an OFDM system with $N$ sub-carriers is considered. At the UAV transmitter, the modulated data symbol frequency-domain (FD) vector ${\mathbf{x}}_{\mathrm{FD}} \in \mathbb{C}^{N \times 1}$ is mapped to the time-domain (TD) vector ${\mathbf{x}}_{\mathrm{TD}} \in \mathbb{C}^{N \times 1}$ by using the inverse discrete fourier transform (IDFT), which is given by
    \begin{equation}\label{EQ1}
        {\mathbf{x}}_{\mathrm{TD}} = {\mathbf{F}}_N^{\mathrm {H}}  \cdot {\mathbf{x}}_{\mathrm{FD}}.
    \end{equation}
By inserting a cyclic prefix (CP), the transmitted signal is given by \cite{r3}
    \begin{equation}\label{EQ2}
        {{\mathbf{s}} = \left[ \begin{array}{l} {{\mathbf{0}}_{L_{\mathrm{CP}} \times (N-L_{\mathrm{CP}})}} \ {{\bf{I}}_{{L_{\mathrm{CP}}}}}\\ \quad\quad\quad\quad {{\mathbf{I}}_N} \end{array} \right] \cdot {{\mathbf{x}}_{\mathrm{TD}}}},
    \end{equation}
where ${\mathbf{s}} \in \mathbb{C}^{(N+L_{\mathrm{CP}}) \times 1}$ with $L_{\mathrm{CP}}$ being the CP length.

From \cite{r1}, $P$ resolvable paths that account for both the LoS and NLoS components are considered for the UAV downlink. The composite vector of channel impulse response (CIR), denoted as $\mathbf{h} \in \mathbb{C}^{P \times 1}$, is represented as \cite{r1}
    \begin{equation}\label{EQ3}
        {\mathbf{h}} = \underbrace {g\sqrt {\frac{k}{{k + 1}}} {\mathbf{\bar c}}}_{ \buildrel \Delta \over = {{\mathbf{h}}_{\mathrm{LoS}}}} + \underbrace {g\sqrt {\frac{1}{{k + 1}}} {\mathbf{\widetilde c}}}_{ \buildrel \Delta \over = {{\bf{h}}_{\mathrm{NLoS}}}},
    \end{equation}
where ${{\mathbf{h}}_{\mathrm{LoS}}} \in \mathbb{C}^{P \times 1}$ and ${{\mathbf{h}}_{\mathrm{NLoS}}} \in \mathbb{C}^{P \times 1}$ stand the CIRs of the LoS and NLoS link, respectively; $g$ denotes the large scale fading factor, $k$ is the Rician factor, ${\mathbf{\bar c}}$ is related to the LoS component, and ${\mathbf{\widetilde c}}$ is related to the NLoS component satisfying ${\mathbf{\widetilde c}} \sim \mathcal{CN}(\mathbf{0}, \mathbf{I}_{P})$.

At the gBS, the received TD signal vector ${\mathbf{r}} \in \mathbb{C}^{(N+L_{\mathrm{CP}}) \times 1}$ is given by
    \begin{equation}\label{EQ5}
        {\mathbf{r}} = \widetilde {\mathbf{h}} \cdot {\mathbf{s}} + {\mathbf{n}},
    \end{equation}
where ${\widetilde {\mathbf{h}}} \in \mathbb{C}^{(N+L_{\mathrm{CP}}) \times (N+L_{\mathrm{CP}})}$ is a Toeplitz circulant matrix formed by $\mathbf{h}$, and $\mathbf{n} \in \mathbb{C}^{(N+L_{\mathrm{CP}}) \times 1}$ denotes the complex additive white Gaussian noise vector satisfying ${\mathbf{n}} \sim \mathcal{CN}(\mathbf{0}, \sigma^2 \mathbf{I}_{N+L_{\mathrm{CP}}})$. From \cite{r2}, by denoting the composite CIR vector as $\mathbf{h}=[h_0, h_1, \cdots,h_p,\cdots, h_{P-1}]^{\mathrm{T}}$, the first row of ${\widetilde {\mathbf{h}}}$ can be expressed as \cite{r2}
    \begin{equation}\label{EQ6}
       {\widetilde {\mathbf{h}}}_{1} = [h_0, \mathbf{0}_{1 \times (N+L_{\mathrm{CP}}-P)}, h_{P-1}, \cdots, h_2, h_1].
    \end{equation}
After removing the CP, the received signal $\mathbf{r}$ is mapped to $\mathbf{y}_\mathrm{TD} \in \mathbb{C}^{N \times 1}$, which is given by \cite{r3}
    \begin{equation}\label{EQ7}
        \mathbf{y}_\mathrm{TD} = \left[ {{\mathbf{0}}_{N \times L_{\mathrm{CP}}} \quad {{\bf{I}}_N}} \right] \cdot \mathbf{r}.
    \end{equation}
By using the discrete fourier transform (DFT), the TD signal vector $\mathbf{y}_\mathrm{TD}$ is mapped to the FD vector $\mathbf{y}_\mathrm{FD} \in \mathbb{C}^{N \times 1}$, which is expressed as
    \begin{equation}\label{EQ8}
        \mathbf{y}_\mathrm{FD} = {\mathbf{F}}_N  \cdot {\mathbf{y}}_{\mathrm{TD}}.
    \end{equation}

With the FD signal vector $\mathbf{y}_\mathrm{FD}$, we exploit the prior information from LoS sensing to improve the CE accuracy of UAV-assisted OFDM systems in Section III.

\vspace{-0.5cm}

\section{LoS Sensing-based Channel Estimation}

In this section, we elaborate on the proposed LoS sensing-based enhanced CE (LoS-EnCE), which leverages path detection thresholds. Section III-A provides an in-depth explanation of the LoS sensing method, while Section III-B focuses on the design of the path detection thresholds. Subsequently, in Section III-C, we present the development of the LoS-EnCE method, incorporating the carefully designed path detection thresholds.

\vspace{-0.5cm}
\subsection{LoS Sensing}

In UAV-assisted OFDM systems, there is a high likelihood of LoS scenarios occurring in the UAV-to-Ground (U2G) links \cite{r4}\cite{r5}. As mentioned in \cite{r6}, the LoS path is considerably stronger, approximately $20 \text{ dB}$ in comparison to the NLoS paths. These observations highlight the common existence of the LoS path in U2G links and its detectability, motivating us to leverage this prior information for enhancing the CE accuracy.

With the CIRs, the kurtosis of the received power, as defined in \cite{r7}, is employed to detect the existence of LoS path. By denoting the kurtosis as $\kappa$, we have \cite{r7}
    \begin{equation}\label{EQ9}
        \kappa  =  \frac{{E\left[ {{{\left( {\left| {h\left( \tau  \right)} \right| - {\mu _{\left| {h\left( \tau  \right)} \right|}}} \right)}^4}} \right]}}{{\sigma _{\left| {h\left( \tau  \right)} \right|}^4}},
    \end{equation}
where $\tau$, $h\left( \tau  \right)$, $\mu _{\left| {h\left( \tau  \right)} \right|}$ and $\sigma _{\left| {h\left( \tau  \right)} \right|}$ denote the tap delay, received CIRs, and the mean and standard deviation of $\left|h\left( \tau  \right)\right|$, respectively. Since the kurtosis in LoS scenarios significantly exceeds that in NLoS scenarios, detecting the presence of LoS path becomes feasible when the computed $\kappa$ is larger than a pre-defined threshold as $\zeta$. To determine the pre-defined threshold, the methodologies outlined in \cite{r12} and \cite{r13} can be referred. For instance, in \cite{r13}, the threshold is set as nearly $50$ for LoS scenarios at a distance of $9.5$ meters. It is important to note that the LoS sensing method in \cite{r7} serves as an illustrative example in this letter. Other LoS sensing methods can also be employed for this work.

\vspace{-0.5cm}
\subsection{LoS-aided Detection Threshold for Resolvable Paths}

With the sensed LoS/NLoS scenarios, we design detection thresholds to detect the resolvable paths. If the NLoS scenarios are detected, an optimized threshold based on the constant false alarm rate (CFAR) is constructed. According to \cite{r8}, the CFAR-based threshold (CFAR-Thr), denoted by ${\widetilde{T}_h}$, is derived as \cite{r10}
    \begin{equation}\label{EQ15}
        {\widetilde{T}_h} = \sqrt { - 2{{\widehat \sigma }^2}\ln {P_f}},
    \end{equation}
where ${{\widehat \sigma }^2}$ denotes the equivalent noise variable by involving $N-L_{\mathrm{CP}}$ CIR samples, and $P_f$ is the pre-defined false alarm probability.

However, this CFAR-Thr ${\widetilde{T}_h}$ does not fully exploit the advantages of the sensed LoS path, rendering it susceptible to multi-path interference and significantly degrading the performance of resolvable paths detection. To tackle this issue, a LoS-aided threshold (LoS-AidThr) is designed by leveraging the LoS prior information comprehensively in this letter.

With the CFAR-Thr $\widetilde{T}_h$, the LoS-AidThr, denoted as ${T}_h$, is given by
    \begin{equation}\label{EQ16}
    {T}_h = \varepsilon  \cdot {\widetilde{T}_h} \cdot \left( {1 + {\delta_{\mathrm{LoS}}}} \right),
    \end{equation}
where $\varepsilon$ denotes the threshold factor. From\cite{epsilon}, $\varepsilon = {{P_f}^{-1/(L^2-1)}-1}$ with ${P_f}$ and $L^2-1$ representing the pre-defined false alarm probability and the size of reference window, respectively. ${\delta_{\mathrm{LoS}}}$ stands the path factor associated with the sensed LoS path. For the estimated CIRs, we describe ${\delta_{\mathrm{LoS}}}$ as
    \begin{equation}\label{EQ17}
        {{\delta }_{\mathrm{LoS}}}=\frac{\left| h_{\mathrm{LoS}} \right|}{\sum\limits_{j=1}^{_{{{L}_{\mathrm{CP}}}/2}}{\left| h\left( {{\tau }_{j}} \right) \right|}}+\delta _{\mathrm{noise}}^{2},
    \end{equation}
where $h_{\mathrm{LoS}}$ stands the complex gain of the LoS path, $h(\tau_j)$ represents the complex gain of the $j$-th path with the path delay being $\tau_j$, and $\delta _{\mathrm{noise}}$ is the noise factor.
Although a LoS path exists, a significant amount of scatter leads to the existence of NLoS paths. Additionally, there is inevitably estimation noise at the receiver. Therefore, in this letter, $\delta _{\mathrm{noise}}$ is defined as
    \begin{equation}\label{EQ18}
       {{\delta }_{\mathrm{noise}}}={\sum\limits_{j={{L}_{\mathrm{CP}}}+1}^{_{N}}{\left| h\left( {{\tau }_{j}} \right) \right|}}{\bigg{/}}{\sum\limits_{j=1}^{_{N}}{\left| h\left( {{\tau }_{j}} \right) \right|}}\;.
    \end{equation}
It is worth noting that the carefully constructed LoS-AidThr ${T}_h$ is designed to characterize the significance of the LoS path.
Additionally, noise effects are also considered, thereby facilitating the identification of resolvable paths.
Subsequently, with the designed detection thresholds, we propose a LoS sensing-based CE scheme in Section III-C to enhance the CE accuracy.

\vspace{-0.5cm}
\subsection{LoS Sensing-based CE Enhancement}
\setlength{\textfloatsep}{0pt}
    \begin{algorithm}[h]
        \SetAlgoLined
        \caption{The algorithm of LoS-EnCE}
        \KwIn{The transmitted signal ${\mathbf{x}}_\mathrm{FD}$, the received signal ${\mathbf{y}}_\mathrm{FD}$, the kurtosis $\kappa$, the pre-defined threshold $\zeta$, CFAR-Thr ${\widetilde{T}_h}$, and LoS-AidThr ${{T}_h}$.}
        \KwOut{The enhanced CIR ${\widehat{\mathbf{h}}_{\mathrm{en}}}$.}
        Estimate the initial feature of CE as ${\widetilde{\mathbf{h}}}_{\mathrm{LS}}$ by using Eq.~(\ref{EQ19}).\\
        Obtain TD ${{\widehat{\mathbf{h}}}_{\mathrm{LS}}}$ transformed from FD ${\widetilde{\mathbf{h}}}_{\mathrm{LS}}$ according to Eq.~(\ref{EQ20}).\\
        \textbf{LoS Sensing:}\\
        Calculate the kurtosis $\kappa$ based on the estimated CIR ${{\widehat{\mathbf{h}}}_{\mathrm{LS}}}$ by using Eq.~(\ref{EQ9}).\\
        \textbf{CE Enhancement:}\\
        \eIf{$\kappa > \zeta$}
        {LoS scenario can be determined, the LoS-AidThr ${{T}_h}$ is employed to obtain the enhanced CIR ${\widehat{\mathbf{h}}_{\mathrm{en}}}$ according to Eq.~(\ref{EQ21}).}
        {NLoS scenario is determined, the CFAR-Thr ${\widetilde{T}_h}$ is applied to obtain the enhanced CIR ${\widehat{\mathbf{h}}_{\mathrm{en}}}$ by using Eq.~(\ref{EQ22}).}
    \end{algorithm}

Based on the CFAR-Thr ${\widetilde{T}_h}$ and the LoS-AidThr ${T}_h$, we develop the LoS-EnCE as follows.
The procedure of LoS-EnCE is summarized in \textbf{Algorithm 1}.

With the received signal ${\mathbf{y}}_{\mathrm{FD}}$ (as provided in Eq.~(\ref{EQ8})), the classic LS estimation in the FD is employed to obtain the initial feature of CE, denoted as ${\widetilde{\mathbf{h}}}_{\mathrm{LS}} \in \mathbb{C}^{N \times 1}$, is expressed by \cite{r14}
    \begin{equation}\label{EQ19}
        {{\widetilde{\mathbf{h}}_{\mathrm{LS}}}}={{\left[ \frac{{{y}_{0}}}{{{x}_{0}}},\frac{{{y}_{1}}}{{{x}_{1}}},\cdots, \frac{{{y}_{N-1}}}{{{x}_{N-1}}} \right]}^{\mathrm{T}}},
    \end{equation}
where $y_i$ and $x_i$ with $i=0,1,\cdots,N-1$ are the $i$-th entries of ${\mathbf{y}}_{\mathrm{FD}}$ and ${\mathbf{x}}_{\mathrm{FD}}$, respectively. Then, the FD ${\widetilde{\mathbf{h}}}_{\mathrm{LS}}$ is transformed to the TD, creating the estimated CIR (denoted as ${\widehat{\mathbf{h}}}_{\mathrm{LS}} \in \mathbb{C}^{N \times 1}$). This transformation is achieved by using an $N$-point IDFT, expressed as
    \begin{equation}\label{EQ20}
        {{\widehat{\mathbf{h}}}_{\mathrm{LS}}}= \mathbf{F}_N^\mathrm{H} \cdot {{{\widetilde{\mathbf{h}}}_{\mathrm{LS}}}}.
    \end{equation}

Subsequently, the LoS sensing method presented in Section III-A is utilized to detect the presence of LoS path, thereby determining whether the current scenario is an LoS or NLoS scenario. Specifically, the estimated CIR ${{\widehat{\mathbf{h}}}_{\mathrm{LS}}}$ is substituted into Eq.~(\ref{EQ9}) to detect the existence of LoS path using the pre-defined threshold in \cite{r13}. Depending on the determined LoS/NLoS scenario, we employ the corresponding LoS/NLoS detection threshold to identify resolvable paths, thereby enhancing the CE accuracy by eliminating potential false paths caused by noise or interference.

As depicted in Fig.~\ref{figure1}, the enhanced CIR ${\widehat{\mathbf{h}}_{\mathrm{en}} \in \mathbb{C}^{N \times 1} }$ is achieved by using the LoS and NLoS detection thresholds. We use ${{\widehat{h}}_{\mathrm{LS},n}}$ and ${{\widehat{h}}_{\mathrm{en},n}}$ with $n=0,1,\cdots,N-1$ to represent the $n$-th entries of ${\widehat{\mathbf{h}}}_{\mathrm{LS}}$ and ${\widehat{\mathbf{h}}_{\mathrm{en}}  }$, respectively. Then, based on the CFAR-Thr ${\widetilde{T}_h}$ and LoS-AidThr ${{T}_h}$, $\left|{{\widehat{h}}_{\mathrm{LS},n}}\right|$ is employed to determine the resolvable paths in both LoS and NLoS scenarios. For the sensed LoS scenarios, we have
    \begin{equation}\label{EQ21}
        \left\{ \begin{array}{l}
        {{\widehat{h}}_{\mathrm{en},n}} = 0, \text{~if~} \left|{{\widehat{h}}_{\mathrm{LS},n}}\right| < {{T}_{h}} \\\\
        {{\widehat{h}}_{\mathrm{en},n}} = {{\widehat{h}}_{\mathrm{LS},n}}, \text{~else~}
        \end{array} \right..
    \end{equation}
Similarly, the $n$-th entry of ${\widehat{\mathbf{h}}_{\mathrm{en}}  }$ in NLoS scenarios is given by
    \begin{equation}\label{EQ22}
        \left\{ \begin{array}{l}
        {{\widehat{h}}_{\mathrm{en},n}} = 0, \text{~if~} \left|{{\widehat{h}}_{\mathrm{LS},n}}\right| < {{\widetilde{T}}_{h}} \\\\
        {{\widehat{h}}_{\mathrm{en},n}} = {{\widehat{h}}_{\mathrm{LS},n}}, \text{~else~}
        \end{array} \right..
    \end{equation}
Thus, according to Eq.~({\ref{EQ21}}) and Eq.~({\ref{EQ22}}), the enhanced CIR ${\widehat{\mathbf{h}}_{\mathrm{en}}}$ is achieved by reserving the CIR taps that surpass its detection threshold.

\section{SIMULATION RESULTS}

    \begin{figure*}[t]
        \hspace{0.5cm}
        \begin{minipage}[t]{0.33\linewidth}
        \centering
        \includegraphics[scale=0.32]{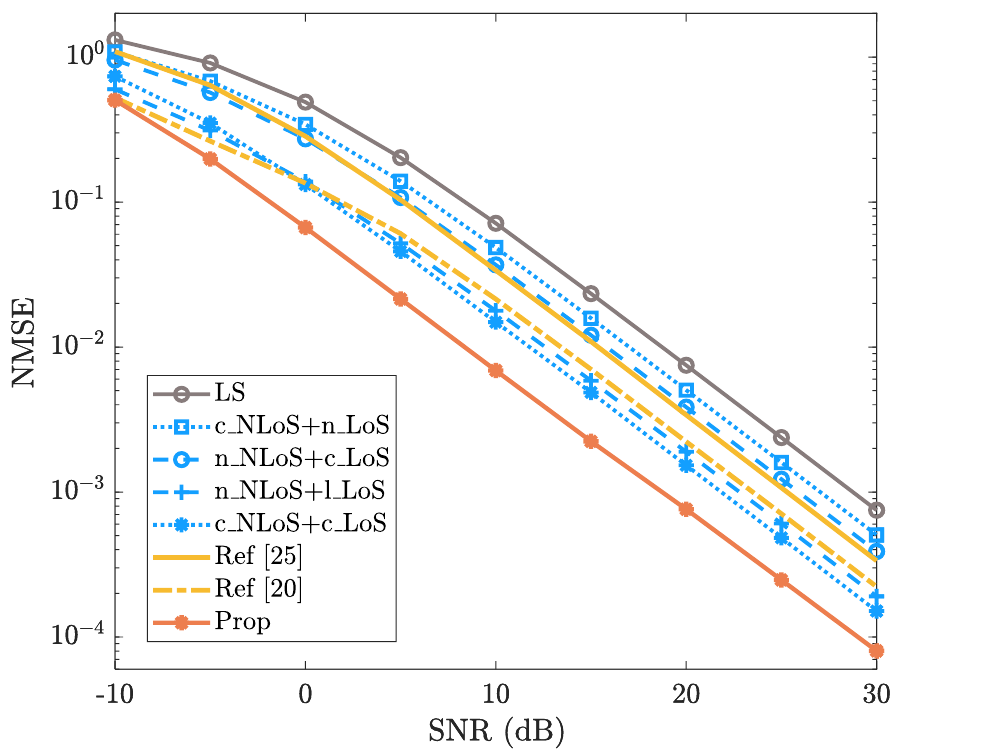}
        \captionsetup{font={footnotesize}}
        \caption{NMSE vs. SNR.}\centering
        \label{figure2}
        \end{minipage}
        \hspace{-0.5cm}
        \begin{minipage}[t]{0.33\linewidth}
        \centering
        \includegraphics[scale=0.32]{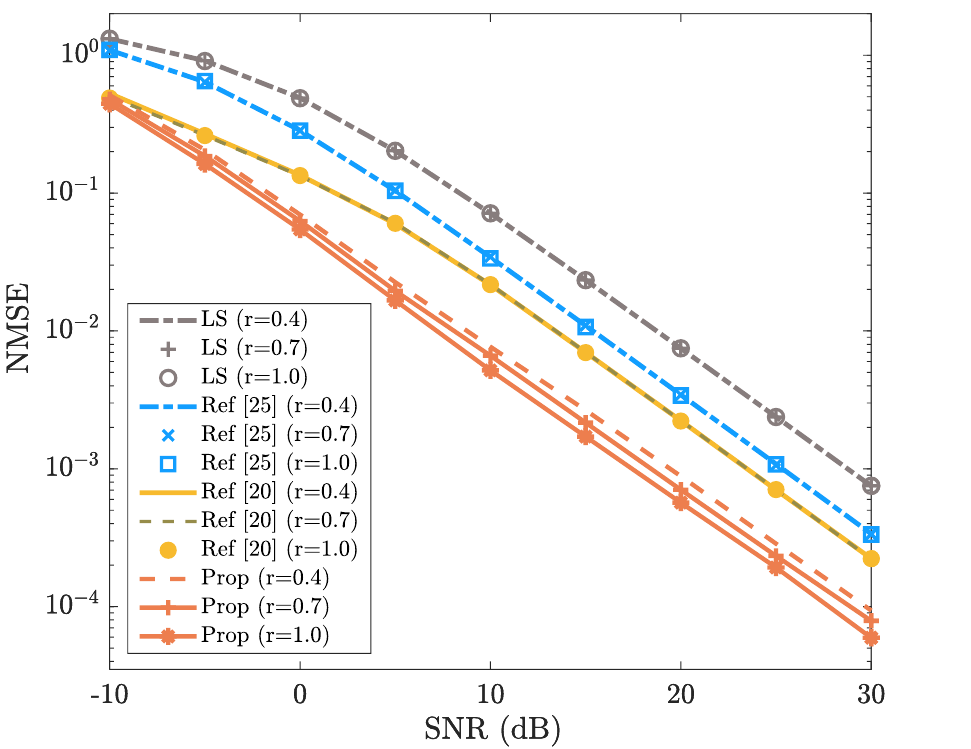}
        \captionsetup{font={footnotesize}}
        \caption{NMSE vs. SNR, where $r=0.4$,\\ $r=0.7$ and $r=1.0$ are considered.}
        \label{figure3}
        \end{minipage}
        \hspace{-0.5cm}
        \begin{minipage}[t]{0.33\linewidth}
        \centering
        \includegraphics[scale=0.32]{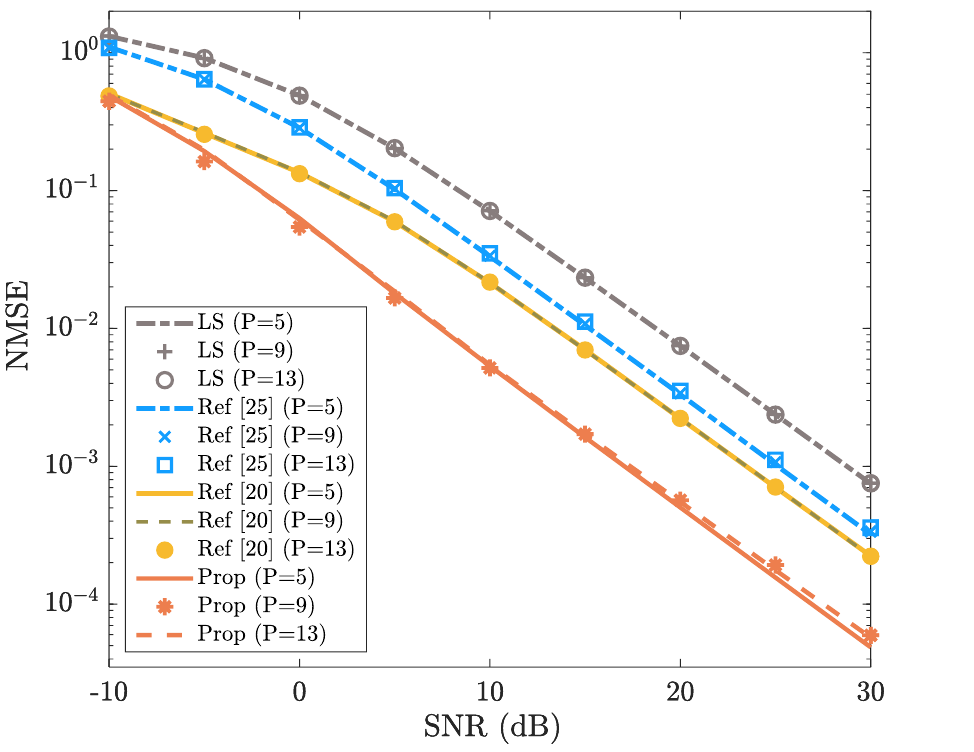}
        \captionsetup{font={footnotesize}}
        \caption{NMSE vs. SNR, where $P=5$,\\ $P=9$ and $P=13$ are considered.}
        \label{figure4}
        \end{minipage}
    \end{figure*}

In this section, we validate the effectiveness and robustness of the proposed CE method by numerical simulations. In Section IV-A, the basic parameters involved in the simulations are given. Then, we verify the effectiveness of the optimized threshold design and the proposed LoS sensing-based CE method in Section IV-B. The robustness analysis against the parameter variations is presented in Section IV-C.

\vspace{-0.5cm}
\subsection{Parameters Settings}

The basic parameters involved in the simulations are as follows. We set $N=512$, $L_{\mathrm{CP}}=64$, and the $4$-quadrature amplitude modulation is employed as the data modulation mode.
For the channel model, we consider the tapped delay line model with $20$ taps. Due to the channel sparsity, $P=10$ is utilized in the simulations.
From \cite{r1}, the large scale fading coefficients are normalized and uniformly distributed over the range $[0.1,1]$. Due to the nature of UAV-assisted communication scenarios, LoS path exists with a high probability. To represent the probability of the LoS path existence, we introduce a scaling factor, denoted as $r$, and set $r=0.8$ in the simulations. In the LoS scenarios, the Rician factor $k$ is randomly generated within the interval $ [3,13]$. In the NLoS scenarios, $k$ is fixed as zero. The pre-defined false alarm probability is set as $P_f = 10^{-3}$\cite{epsilon}\cite{r15}. From\cite{epsilon}, the threshold factor $\varepsilon$ is set to $0.6$ with $P_f = 10^{-3}$ and $L=4$. The SNR is defined as $\mathrm{SNR}={p_t}/{\sigma}^2$ \cite{r1}. The normalized mean square error (NMSE) is used to evaluate the performance of the proposed CE method, which is defined as \cite{r1}
    \begin{equation}\label{EQ23}
        \mathrm{NMSE} = \frac{1}{N}\sum\limits_{n = 0}^{N - 1} {\frac{{{{\left\| {{{\widehat h}_{en,n}} - {h_n}} \right\|}_2^2}}}{{{{\left\| {{h_n}} \right\|}_2^2}}}},
    \end{equation}
where $\mathbf{h}$ in Eq.~(\ref{EQ3}) is extended to an $N$-length vector by padding zeros to form $h_n$ with $n=0,\cdots,{ N}-1$.
    \setlength{\textfloatsep}{0pt}
    \newcommand{\tabincell}[2]{\begin{tabular}{@{}#1@{}}#2\end{tabular}}
    \begin{table}[ht]
    \captionsetup{font={footnotesize}}
    \caption{The abbreviations of the methods used in the simulations.}\centering
    \label{table1}
    \scalebox{0.63}{
        \begin{tabular}{|c|c|}
        \hline
        Methods & Descriptions \\
        \hline
        LS & LS CE \\
        \hline
        Prop & \tabincell{c}{proposed method} \\
        \hline
        Ref\cite{r8} & \tabincell{c}{threshold-based CE enhancement discussed in \cite{r8}} \\
        \hline
        Ref\cite{r16} & \tabincell{c}{threshold-based CE enhancement discussed in \cite{r16}} \\
        \hline
        n\_NLoS+c\_LoS & \tabincell{c}{absence of the CFAR-Thr ${\widetilde{T}_h}$ for the NLoS scenarios, LoS scenarios\\ not employs the LoS-AidThr ${{T}_h}$} \\
        \hline
        n\_NLoS+l\_LoS & \tabincell{c}{absence of the CFAR-Thr ${\widetilde{T}_h}$ for the NLoS scenarios, LoS scenarios \\ employs the LoS-AidThr ${{T}_h}$} \\
        \hline
        c\_NLoS+n\_LoS & \tabincell{c}{CFAR-Thr ${\widetilde{T}_h}$ is used in the NLoS scenarios, while LoS-AidThr \\ ${{T}_h}$ is not applied for the sensed LoS scenarios} \\
        \hline
        c\_NLoS+c\_LoS & \tabincell{c}{CFAR-Thr ${\widetilde{T}_h}$ is used in both the NLoS and LoS scenarios} \\
        \hline
        \end{tabular}
    }
    \end{table}

For expression simplicity, the abbreviations of the methods used in the simulations are summarized in TABLE \ref{table1}.

\vspace{-0.5cm}
\subsection{Effectiveness Analysis}

To validate the effectiveness of the proposed method, Fig.~\ref{figure2} plots the NMSE performance. From Fig.~\ref{figure2}, the NMSE of ``Prop'' is lower than those of ``LS'', ``Ref\cite{r8}'', and ``Ref\cite{r16}'' in all given SNR region. For example, when $\mathrm{SNR} = 20$ dB, the ``Prop'' achieves a NMSE of $7 \times 10^{-4}$, while the NMSEs of ``LS'', ``Ref\cite{r8}'', and ``Ref\cite{r16}'' are all larger than $10^{-3}$. This indicates that the ``Prop'' enhances the NMSE performance of ``LS'', and is more effective than ``Ref\cite{r8}'' and ``Ref\cite{r16}''. The reason is that the proposed method makes full advantage of LoS prior information to design the optimized threshold LoS-AidThr ${{T}_h}$. In conclusion, the proposed method effectively improves the CE accuracy of ``LS'', ``Ref\cite{r8}'', and ``Ref\cite{r16}'' by leveraging the sensed LoS information.

Without LoS sensing, both the NLoS and LoS scenarios adopt CFAR-Thr ${\widetilde{T}_h}$ as the detection threshold. From Fig.~\ref{figure2}, the NMSE of ``Prop'' is lower than that of ``c\_NLoS+c\_LoS'' for each given SNR. This validates the effectiveness of the LoS sensing. Specifically, the proposed method effectively enhances the NMSE performance according to the sensed LoS/NLoS scenario and the associated path detection threshold, i.e., LoS-AidThr ${{T}_h}$. Thus, the LoS sensing is effective in enhancing the CE accuracy.

Fig.~\ref{figure2} also shows the effectiveness of LoS-AidThr ${{T}_h}$ and CFAR-Thr ${\widetilde{T}_h}$, respectively. The NMSE of ``n\_NLoS+l\_LoS'' is lower than that of ``n\_NLoS+c\_LoS'' for each given SNR. This reflects that the LoS-AidThr ${{T}_h}$ is more effective than CFAR-Thr ${\widetilde{T}_h}$ in enhancing the CE accuracy for the LoS branch by detecting resolvable paths. In addition, the NMSE of ``c\_NLoS+n\_LoS'' is lower than that of ``LS''. This verifies the effectiveness of CFAR-Thr ${\widetilde{T}_h}$ in improving the CE accuracy for the NLoS branch.

On the whole, the proposed method presents better NMSE performance compared with ``LS'', ``Ref\cite{r8}'', and ``Ref\cite{r16}'', and both the LoS sensing and the optimized detection thresholds prove effective in enhancing the CE accuracy.

\vspace{-0.4cm}
\subsection{Robustness Analysis}

In this subsection, we discuss the robustness of the proposed method against the impacts of $r$ and $P$. With the exception of the parameters discussed in the robustness analysis, other basic parameters remain the same as those detailed in Section IV-A.

\subsubsection{Robustness against $r$ }

In Fig.~{\ref{figure3}}, we illustrate the robustness of the proposed method against the impact of parameter $r$, where $r=0.4$, $r=0.7$, and $r=1.0$ are considered. From Fig.~{\ref{figure3}}, the NMSE of ``Prop'' decreases as $r$ increases. For example, when $\mathrm{SNR}=30$ dB, the NMSE of ``Prop'' with $r=0.4$ is $9.4 \times 10^{-5}$, while its NMSEs decrease to  $7.8 \times 10^{-5}$ and $5.9 \times 10^{-5}$ for the cases where $r=0.7$ and $r=1.0$, respectively. This improvement benefits from the advantages of the sensed LoS prior information. Moreover, Fig.~{\ref{figure3}} shows that the ``Prop'' achieves a lower NMSE than those of ``LS'', ``Ref\cite{r8}'', and ``Ref\cite{r16}'' for each given value of $r$. When $r=0.7$, the ``Prop'' achieves approximately 12 dB NMSE improvement over the classic ``LS'', and surpasses the NMSE gains of both ``Ref\cite{r8}'' and ``Ref\cite{r16}'' by more than 5 dB. Thus, against the variations of $r$, the proposed method outperforms ``LS'', ``Ref\cite{r8}'', and ``Ref\cite{r16}'' in terms of the NMSE performance.

\subsubsection{Robustness against $P$}

Fig.~{\ref{figure4}} presents the robustness of the proposed method with different resolvable paths, i.e., $P=5$, $P=9$, and $P=13$ are considered. The ``Prop'' exhibits similar NMSEs against the variations of resolvable paths. For example, when $\mathrm{SNR}=20$ dB, the NMSEs of ``Prop'' with different values of $P$ are all about $5.5 \times 10^{-4}$. Furthermore, from Fig.~{\ref{figure4}}, it can be seen that the ``Prop'' achieves a lower NMSE than those of ``LS'', ``Ref\cite{r8}'', and ``Ref\cite{r16}'' for each given SNR. As $\mathrm{SNR}=10$ dB, the NMSE of ``Prop'' is about $5 \times 10^{-3}$, while the NMSEs of ``LS'', ``Ref\cite{r8}'', and ``Ref\cite{r16}'' are all larger than $2 \times 10^{-2}$. This illuminates that the NMSE performance of ``Prop'' outperforms those of ``LS'', ``Ref\cite{r8}'', and ``Ref\cite{r16}'' against the variations of resolvable paths. On the whole, against the impact of $P$, the proposed method improves the NMSEs of ``LS'', ``Ref\cite{r8}'', and ``Ref\cite{r16}''.

\vspace{-0.4cm}
\section{Conclusion}

In this letter, we have investigated a LoS sensing-based CE method to enhance the CE accuracy in UAV-assisted OFDM systems. Based on the fact that LoS path has a high occurrence probability in the U2G wireless scenarios of UAV communication systems, and inspired by ISAC ideas, the letter first develops a LoS sensing method to detect the LoS path existence. According to the acquired LoS/NLoS path prior information, we design detection thresholds to capture resolvable paths in LoS and NLoS scenarios, respectively. By utilizing the designed detection thresholds, the denoising processing is applied to classical LS CE, thereby enhancing estimation accuracy. Experimental results validate that the proposed method effectively improves the CE accuracy and exhibits its robustness to the parameter variations.
\vspace{-0.2cm}

\nocite{*}

\bibliographystyle{ieeetran}

\appendices
\ifCLASSOPTIONcaptionsoff
  \newpage
\fi
\end{document}